\documentclass[fleqn,usenatbib]{mnras}

\usepackage{newtxtext,newtxmath}

\usepackage[T1]{fontenc}

\DeclareRobustCommand{\VAN}[3]{#2}
\let\VANthebibliography\thebibliography
\def\thebibliography{\DeclareRobustCommand{\VAN}[3]{##3}\VANthebibliography}

\usepackage{graphicx}	
\usepackage{amsmath}	
\usepackage{hyperref}

\newcommand{\sm}{\textsubscript{\(\odot\)}}

\newcommand{\bvf}{Brunt--V\"ais\"al\"a}

\newcommand{\msol}{\ensuremath{\mathrm{M}_\odot} }

\title[B-Field Topology in Core-Convecting Stars]{Dynamo Simulations Confirm Predominantly Toroidal Fields in Near-Core Region of an Intermediate-Mass Star}

\author[Ratnasingam \& Rogers]{
R.P. Ratnasingam,$^{1}$\thanks{E-mail: rathishprevin@gmail.com}
T.M. Rogers,$^{1,2}$\thanks{E-mail: tamara.rogers@newcastle.ac.uk}
\\
$^{1}$School of Mathematics, Statistics and Physics, Newcastle University, Newcastle upon Tyne, NE1 7RU, UK\\
$^{2}$Planetary Science Institute, Tucson, AZ 85721, USA\\
}

\date{Accepted XXX. Received YYY; in original form ZZZ}

\pubyear{\the\year{}}

\begin{document}
\label{firstpage}
\pagerange{\pageref{firstpage}--\pageref{lastpage}}
\maketitle

\begin{abstract}
Recent asteroseismic analysis of the main-sequence F star KIC 9244992 has revealed evidence for a strong, predominantly toroidal magnetic field in its deep radiative interior. Motivated by this result, we present three-dimensional anelastic magnetohydrodynamic simulations of a rotating 2 \msol{} main-sequence star to investigate whether such internal magnetic field strengths and geometries naturally arise in global stellar dynamo simulations for stars within a similar mass range. These simulations self-consistently generate magnetic fields with azimuthal components that dominate over radial components by factors of several near the peak of the \bvf{} frequency left behind as the core recedes. Moreover, these field strengths arise despite near-uniform, spherically-averaged, rotation from the near-core region to the surface. When the Reynolds number is greater than $\sim$100 and the appropriate Rossby number is used, the resulting field strengths and geometrical properties are consistent with those inferred asteroseismically for KIC 9244992. These results indicate that a dominant toroidal field in the near-core region is a generic field configuration in core-convecting stars and should be considered in future asteroseismic inferences. 
\end{abstract}

\begin{keywords}
stars: magnetic field -- MHD -- stars: interiors
\end{keywords}



\section{Introduction} \label{sec:intro}
Surveys of hot, dwarf stars (O, B, and A spectral classes) have shown that only 8 -- 10 \% of them possess large-scale detectable magnetic fields on their surfaces \citep{Grunhut2012,Petit2026a}, which are generally explained by fossil fields or dynamo mechanisms. However, magnetic fields are actually expected to be present in all stars \citep{DonatiLandstreet2009}, but in most cases, those fields are either too weak for detection or confined in the deeper interior. 

Asteroseismology now enables the indirect inference of internal properties of stars including magnetic field strengths \citep{Fuller2015,Stello2016,Li2022} and recently such methods have been applied to main-sequence stars \citep{Lecoanet2022a,Vandersnickt2025}. However, in all these detections, the field geometry is typically assumed to be predominantly radial everywhere within the star, contrary to results from dynamo simulations \citep{Brun2005,Browning2008,Featherstone2009,Auguston2016,ratnasingam2024}. More recently, \cite{masao2025} showed that using asymmetries caused by the presence of a magnetic field combined with rotational splitting, the minimum values for both the azimuthal and radial magnetic fields can be inferred deep in the stellar interior for the star KIC 9244992, observed with the Kepler Space Telescope \citep{Borucki2010}. Using 17 g-modes, the minimum values of the azimuthal and radial components of the magnetic field were inferred to be 92 $\pm$ 7 kG and 3.5 $\pm$ 0.1 kG respectively, making this the first work to constrain the poloidal {\it and} toroidal field strengths in the interior of a star. Although the work provided the radial location of these magnetic fields to be less than 50\% the total stellar radius, particular attention was given to the near-core region, defined to be the radius just outside the convective-radiative boundary, within the region of large chemical gradient left behind by a receding core. In a previous work on the same star, \cite{Saio2015} showed that the near-core to near-surface differential rotation was remarkably flat with only a 4\% difference. This, ultimately, raises the question of how a strong toroidal field can be maintained in the presence of weak differential rotation. 

Prior to the above asteroseismic inference, \cite{ratnasingam2024} showed numerically that for a 7 \msol{} star at central hydrogen mass fraction of 0.35, a near-core shear layer was responsible for the amplification of the azimuthal component of the magnetic field in that region. \cite{ratnasingam2024} served to numerically inform the stellar astrophysics community that the azimuthal component of the magnetic field should dominate over the radial component of the magnetic field in the near-core region for core--convecting stars. With the more recent observational work constraining field strengths in \cite{masao2025}, we follow up on \cite{ratnasingam2024}, by performing 3D MHD simulations of a 2 \msol{} mid-main sequence star. In addition to investigating the field strength and geometry, we also present the differential rotation to demonstrate how such a field geometry could develop.  Similar to \cite{ratnasingam2024}, this work builds on previous works such as \cite{Featherstone2009} and \cite{Auguston2016}, which also presented results from numerical simulations of core--convecting stars. Although we adopt the same stellar mass as \cite{Featherstone2009}, our focus is motivated directly by asteroseismic constraints, with particular emphasis on the near-core region marked by the \bvf{} frequency peak. 

\vspace{-0.5cm}
\section{Numerical setup and Evolution} \label{sec:Numerical_setup}
Our 3D spherical simulations are done using {\sc SPIN} \citep{philipp3dpaper}, a pseudospectral code which solves the MHD equations in the anelastic approximation. Discretisation is performed with a finite-difference scheme in the radial direction and spherical harmonics in the horizontal directions. The whole simulation domain is resolved with 1500 radial ($r$) grid points, 128 grid points in the polar ($\theta$) direction and 256 grids in the azimuthal ($\phi$) direction, equivalent to a spherical harmonic $\ell_{\rm max}=84$. The code implements parallel frameworks using Message Passing Interface (MPI) and solves the following MHD equations:
\begin{align}
    \overline{\rho}\frac{\partial \boldsymbol{v}}{\partial t} +\overline{\rho}(\boldsymbol{v}\cdot\boldsymbol{\nabla}\boldsymbol{v})  
    -2\overline{\rho}\boldsymbol{v}\times\boldsymbol{\Omega}   &=\; \nonumber
    \\-\overline{\rho}\overline{g}\,\boldsymbol{\hat{r}} 
    +\boldsymbol{\nabla}\left(\frac{P}{\overline{\rho}}\right) \label{eq:anelastic-momentum} \nonumber
    \\ +\nabla\cdot(2\overline{\rho}\overline{\nu}(e_{ij}-\frac{1}{3}(\nabla\cdot\boldsymbol{v})\delta_{ij})) \nonumber
    \\+\frac{1}{\mu_0}\left(\boldsymbol{\nabla}\times\boldsymbol{B}\right)\times\boldsymbol{B}, 
    \\
    \frac{\partial T}{\partial t} + (\boldsymbol{v}\cdot\boldsymbol{\nabla})T =\; \nonumber
    \\ -v_r\left(\frac{\partial \overline{T}}{\partial r} - (\gamma -1)\overline{T}h_{\rho}\right) + (\gamma - 1) T h_{\rho} v_r  \nonumber
     \\ + \gamma \overline{\kappa}\left(\nabla^2T + (h_{\rho} + h_{\kappa})\frac{\partial T}{\partial r}\right)
    +\frac{\overline{\eta}}{\mu_0 \overline{\rho} \overline{c_p}}\left[\boldsymbol{\nabla}\times\boldsymbol{B}\right]^2,  \label{eq:anelastic-energy}\\ 
    \frac{\partial \boldsymbol{B}}{\partial t} =\; \boldsymbol{\nabla}\times\left[\,\boldsymbol{v}\times\boldsymbol{B}-\overline{\eta}\boldsymbol{\nabla}\times\boldsymbol{B}\,\right], \\
    \boldsymbol{\nabla}\cdot\left[\overline{\rho}\,\boldsymbol{v}\right] =\; 0, \\
    \boldsymbol{\nabla}\cdot\boldsymbol{B} =\; 0 \label{eq:mag-monopole}. 
\end{align}
The radial velocity, $v_r$, together with the latitudinal velocity, $v_{\theta}$, and the azimuthal velocity, $v_{\phi}$, form the velocity vector, $\boldsymbol{v}$. The radial, polar and azimuthal magnetic fields, represented by $B_r$, $B_{\theta}$ and $B_{\phi}$ respectively, form the magnetic field vector, $\boldsymbol{B}$. $T$ and $P$ are the perturbation temperature and pressure, respectively. The quantities with an overline are the radially dependent reference state values density, $\overline{\rho}$, specific pressure heat capacity, $\overline{c_P}$, gravity, $\overline{g}$, temperature, $\overline{T}$, thermal diffusivity, $\overline{\kappa}$, magnetic diffusivity, $\overline{\eta}$, and viscosity, $\overline{\nu}$. The adiabatic index is represented by $\gamma$, which is set to 5/3 and the magnetic permeability, $\mu_0$ is set as the $4\pi$ in cgs units. The inverse density and thermal diffusivity scale heights are written with the symbols, $h_{\rho}$ and $h_{\kappa}$, respectively. 

The reference input state data are taken from the 1D stellar evolution code Modules for Stellar Astrophysics (MESA r23.05.01; \citealt{mesa_1,mesa_2,mesa_3,mesa_4,mesa_5}) for an input model of a 2~M\sm{} star at X$_c$ = 0.5. For the MESA simulations, we set the stellar metallicity, $Z$, to be the solar value of $Z$ = 0.02, the mixing-length parameter to be 1.8 and the convective overshoot profile is set to exponential. We note that this model is effectively "boosted" and convective velocities are approximately 10 $\times$ mixing length theory velocities (MLT). Lastly, the stellar rotation rate, $\Omega$, within our MHD simulation was initially set to a solid body rotation rate of $1.0 \times 10^{-5}$ rad s$^{-1}$ (i.e. 7.3~d), approximately 10 times faster than KIC 9244992.  This faster rotation compensates faster convective velocities in order to achieve a Rossby number that is equivalent to KIC 9244992 (see below).

\begin{figure}
        \centering
        \includegraphics[trim={0.0cm 0.0cm 0 0.0cm},clip,width=\columnwidth]{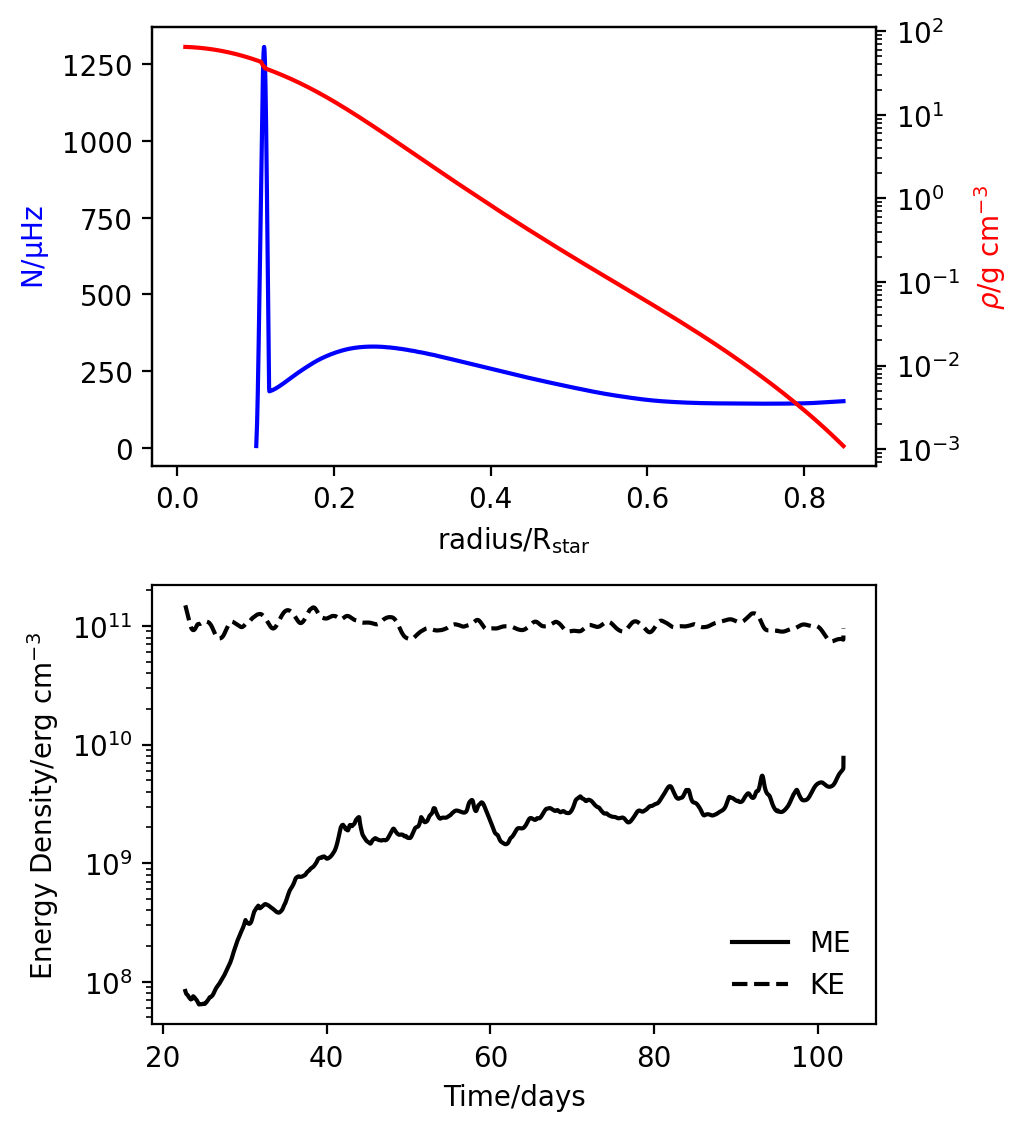}
        \caption{Reference state density, $\overline{\rho}$ (red) and reference state \bvf{} frequency (blue) as a function of radius in units of R$_{\mathrm{star}}$ in the top panel and various energies averaged over the convection zone in the bottom panel. \label{fig:stellar_param}}
        \centering
\end{figure}

We show the smoothed \bvf{} frequency and density profiles of the reference state model as blue and red lines in Fig.~\ref{fig:stellar_param} respectively. We employ similar methods of interpolation and smoothing as those discussed in \cite{ratnasingam2024}. The top of the simulation domain is set to be at 85\% the total stellar radius to ensure numerical stability. We use a non-uniform finite-difference discretisation in the radial direction, where the radial resolution is set to be three times higher in the convection zone compared to the radiation zone. Furthermore, we extend the resolution set in the convection zone into the radiation zone (up to 0.2 R$_{\mathrm{star}}$), to resolve the core-radiative boundary (CRB) better.  

The setup of the simulation begins with a purely hydrodynamical model, which is allowed to evolve in time. The boundary conditions are set to be impermeable and stress-free. The diffusivities are set to be similar to those in \cite{ratnasingam2024}. Once a steady-state evolution is reached, we impose a dipole magnetic field, with the radial field decreasing with the cube of inverse radius ($\propto r^{-3}$). The dipole field strength at the bottom boundary is set to be 100 kG. The magnetic boundary conditions are set as perfect conductor in the inner boundary and potential field matching in the outer boundary. As shown in the bottom panel of Fig.~\ref{fig:stellar_param}, we track the shell-averaged magnetic energy evolution in the convection zone until it reaches a statistically steady state after 40~d of physical time. All analysis presented here uses data from days 40–-100.  

The dominant controlling factors in determining the magnetic field strength in a dynamo simulation are the hydrodynamic ($Re=UL/\nu$) and magnetic ($Rm=UL/\eta$) Reynolds numbers, the magnetic Prandtl number ($Pm=\nu/\eta$) and, crucially, the Rossby number ($U/\Omega L$), where $U$ is the convective velocity and L is the length scale of the system.  As is the case in all astrophysical fluid dynamics, the Reynolds numbers in stars are extremely large, in excess of $10^{10}$ \citep{BRANDENBURG20051}, and numerical simulations can not reproduce this large dynamic range.  It is generally expected that once the Reynolds and magnetic Reynolds numbers are sufficiently large such that inertial and inductive effects dominate over viscous and resistive dissipation, the system enters an asymptotic regime in which the large-scale dynamics become effectively independent of the microscopic diffusivities \citep{Kapya2023}. The simulation we present below has a $Re$ of 240 ($Rm$=960), while the most well resolved global dynamo simulations have Re$\sim$ 300-2000 \citep{Brun2004,BRANDENBURG20051,Brown2011,Auguston2016}. For stellar dynamos the Rossby number also determines the field strength since it is rotation that induces the toroidal field in a typical $\alpha-\Omega$ dynamo and it is generally expected that low Rossby numbers lead to larger magnetic energies (relative to kinetic energy) \citep{Auguston2016,Auguston2019}. KIC 9244992 has a Rossby number of approximately 0.1--0.3, calculated from mixing-length theory velocities, and it is expected that higher mass stars have $Pm$ larger than 1. While our simulation has a rotation rate 10 times that of KIC 9244992, our convective velocities are also approximately 10 times MLT velocities, hence our Rossby number is similar to that of KIC 9244992.  Within the convective core our magnetic Prandtl number $\sim$ 4 and hence, the simulation is in the regime expected for core-convecting stars.  We ran three different simulations varying the Ro and Re of the simulations and an additional simulation with a different initial field strength. In the following we present our most realistic simulation with the largest $Re$ and $Rm$ of 240 and 960 respectively, and Rossby number $\sim 0.1-0.3$, similar to KIC 9244992.  We discuss the other simulations in Section~\ref{sec:Simulation_Parameters}.

Apart from \cite{ratnasingam2024}, there are two other works that this work builds upon, which are \cite{Featherstone2009} and \cite{Auguston2016}. \cite{Auguston2016} presents dynamo simulations of a 10 \msol{} (B-type) star, up to 60\% of the stellar radius, and explores the dependence of convective and magnetic energetics on rotation rate, including variations in proximity to equipartition. \cite{Featherstone2009} present dynamo simulations of a 2 \msol{} star over the inner $\sim$30\% of the stellar radius. They study a variety of initial field configurations and report equipartition and even super-equipartition field strengths within the convection zone reaching 80kG.  While both of these studies investigate field geometries and strength within the convection zone, there is little or no attention given to the near-core region, where asteroseismology is sensitive.  This work addresses that particular problem. 

\section{Magnetic Field Geometry}\label{sec:mag_field_geo}
The main goal of this paper is to present results on the magnetic field topology and strengths from MHD simulations of 2 \msol{} star at X$_c$ = 0.5, following up on findings presented in \citep{ratnasingam2024}, which showed that the magnetic field topology in the near-core region is dominated by the toroidal component of the field. The model presented here differs from the model used in \cite{ratnasingam2024}, as we simulate an F-type star of 2 \msol{}, which is closer to the mass of the star reported in \cite{masao2025}. 

\begin{figure}
        \centering
        \includegraphics[trim={0.cm 0.cm 0.cm 0.0cm},clip,width=\columnwidth]{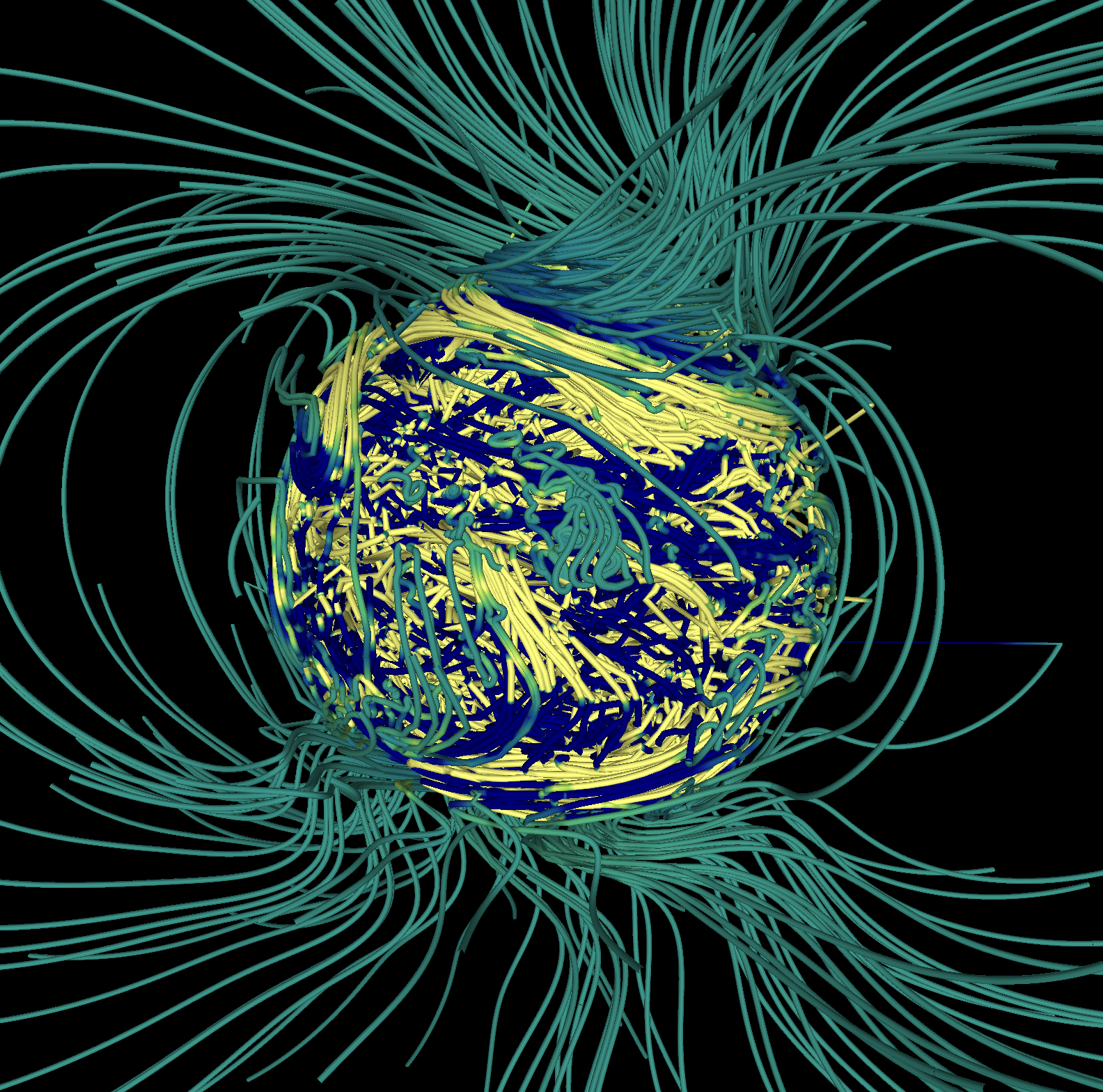}
        \caption{Streamline diagram showing the shape of magnetic field lines in the convection zone and radiation zone after 40~d of physical time. \label{fig:field_streamline}}
        \centering
\end{figure}

Figure~\ref{fig:field_streamline} shows the streamline plot of the magnetic field lines after 40~d of physical time, just as the simulation reaches a steady-state configuration. As mentioned in Section~\ref{sec:Numerical_setup}, the initial seed field is a dipole field, which forms the field lines outside the core (green lines). Within the core, the magnetic field configuration induced by the turbulent convection is more complex, as shown by the yellow and blue lines. Our main focus is the near-core region, just above the convective-radiative boundary (CRB). Within this region, previous works \citep{philipp3dpaper,vanon2023} have shown that layers of strong differential rotation (i.e. shear layers) tend to form. \cite{ratnasingam2024} showed that these shear layers facilitate the induction of the toroidal component of the magnetic field from the poloidal component through the $\Omega$-effect. Our 2 \msol{} MHD model exhibits a significant shear layer at the core-envelope boundary, which is sufficient to drag radial fields into the azimuthal direction, as shown near the polar regions in Fig.~\ref{fig:field_streamline}.    

\begin{figure}
        \centering
        \includegraphics[trim={0.0cm 0.cm 0.cm 0.cm},clip,width=\columnwidth]{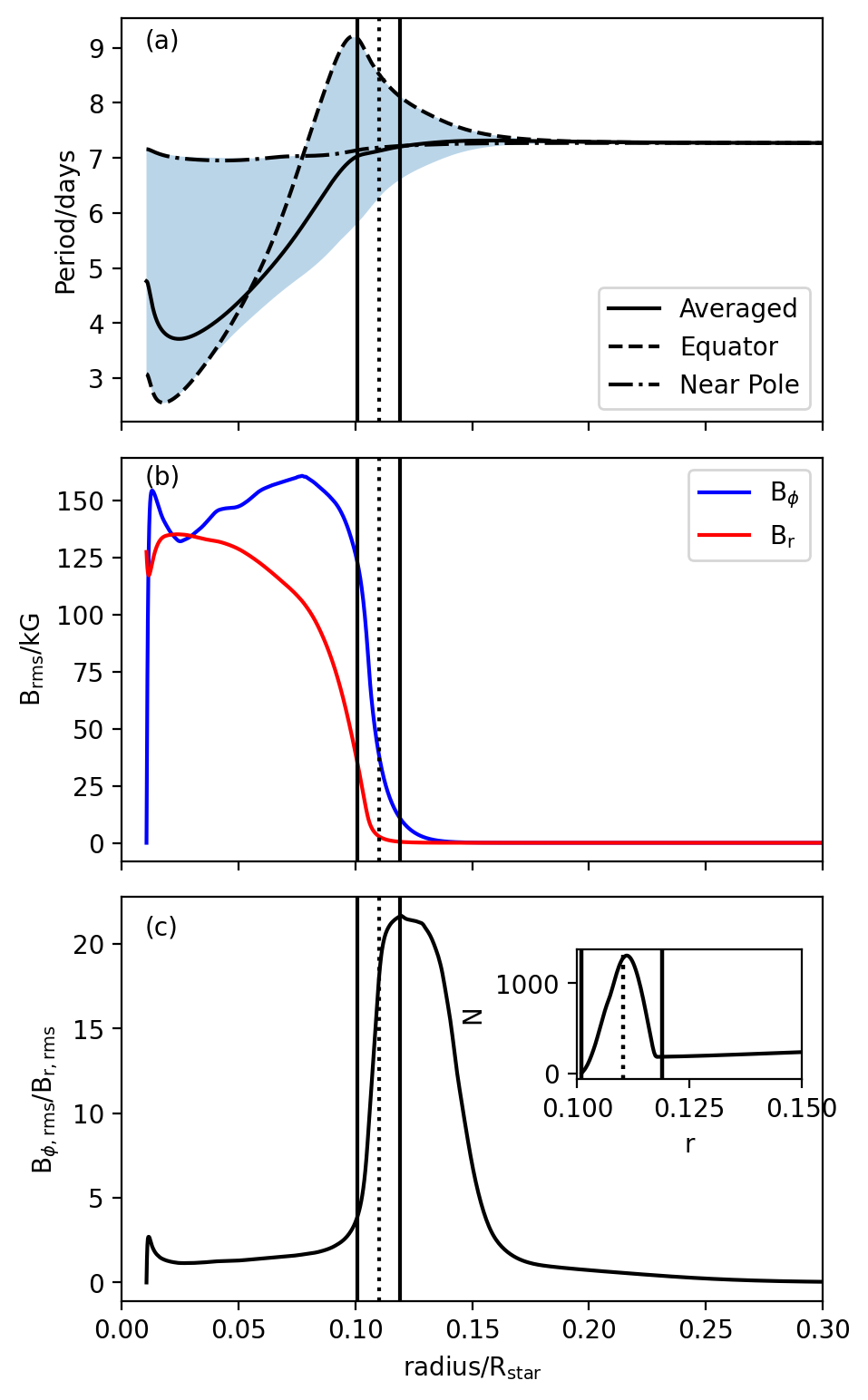}
        \caption{The rotation period (a), root-mean-squared azimuthal and radial magnetic fields (b) and the ratio of root-mean-squared azimuthal to radial magnetic field (c) as a function of radius in units of $R_{\mathrm{star}}$, all within the 30\% of the total stellar radius. The inset in the bottom panel shows the \bvf{} frequency profile within the same region, and the three vertical lines indicate the convective-radiative boundary, the peak of the \bvf{} frequency spike and the base of the \bvf{} frequency spike. All values are time-, latitude- and longitude-averaged. \label{fig:vp-br-bp-r}}
        \centering
\end{figure}

At first glance, this strong near--core differential rotation may seem in contradiction to the asteroseismic inference from \cite{Saio2015}.  In Fig.~\ref{fig:vp-br-bp-r}(a), we show the radial differential rotation profile for the inner region of the simulation as the outer layers show virtually no differential rotation. The solid vertical lines show the extent of the \bvf{} frequency spike, whilst the dotted vertical line marks the location of the maximum \bvf{} frequency. The solid, black curve shows the rotation profile that is averaged over time, latitude and longitude, whilst the region shaded in blue indicates variation in latitude. The rotation profiles near the pole and equator, represented by the dotted-dashed and dashed curves respectively, mark the limits of the profiles over latitude. In general, the rotation profile shows that the convective core rotates at a faster rate than the radiation zone. This is true at almost all latitudes and the maximum rotation rate is more than double the solid-body rotation rate imposed on the model. However, at the CRB, the differential rotation is significantly reduced, reverting to the imposed uniform rotation just above the \bvf{} frequency spike. The important point to highlight here with regard to asteroseismic inferences, is that although there is differential rotation between the convective core and the radiative envelope, the {\it spherically-averaged} differential rotation between the {\it near-core} region, where g-modes probe, and the envelope is very small. Moreover, within the near-core region, there is significant latitudinal differential rotation, which can locally induce a strong toroidal magnetic field, but on {\it average}, appears as uniform rotation.


\begin{figure} 
        \centering
        \includegraphics[trim={0.0cm 0.cm 0.cm 0.cm},clip,width=\columnwidth]{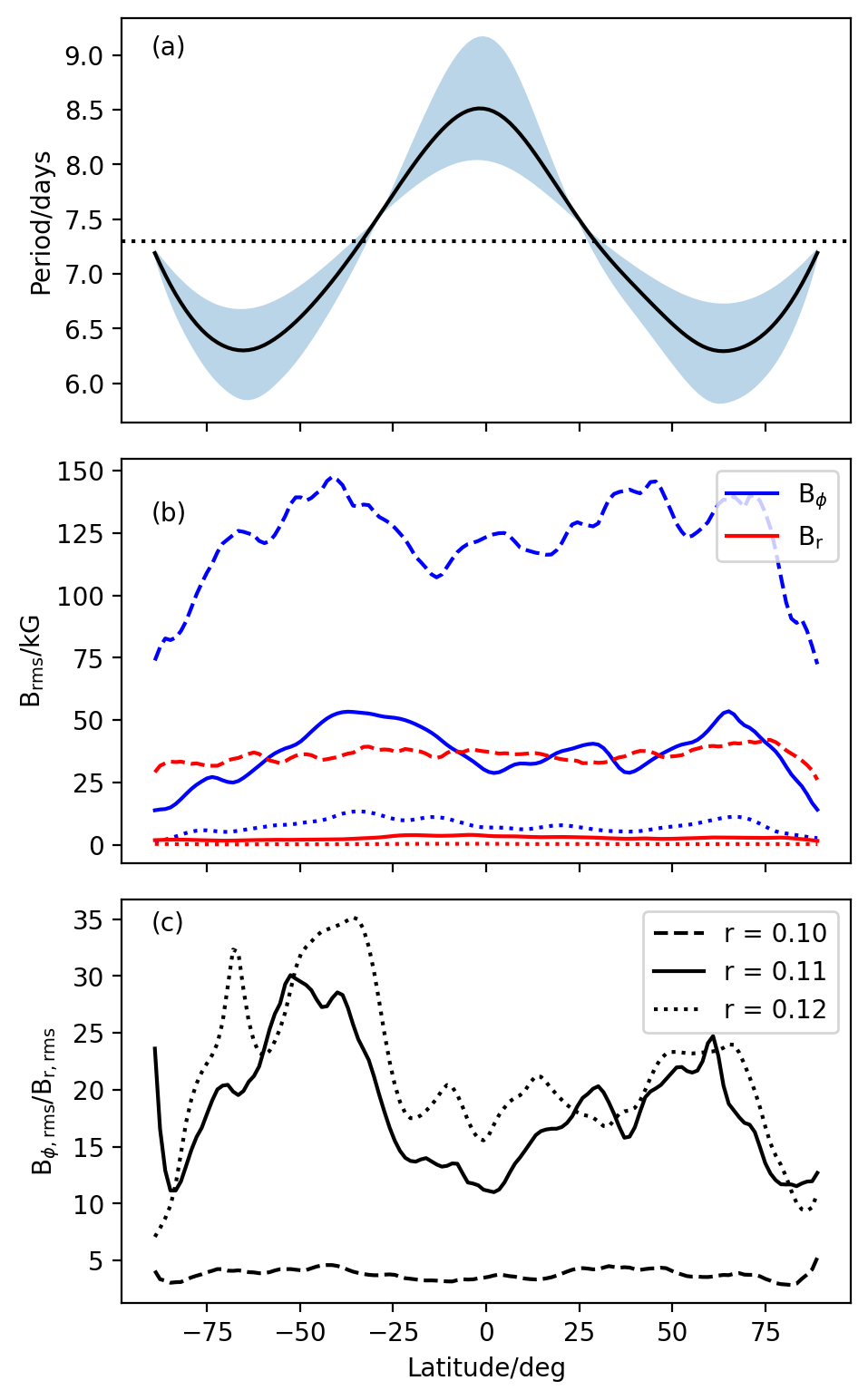}
        \caption{The rotation period (a), root-mean-squared azimuthal and radial magnetic fields (b) and the ratio of root-mean-squared azimuthal to radial magnetic field (c) as a function of latitude. The three radii shown with different linestyles refer to the same radii shown in \ref{fig:vp-br-bp-r}. All values are time- and longitude-averaged. \label{fig:vp-br-bp-r-latdep}}
        \centering
\end{figure}

Figure~\ref{fig:vp-br-bp-r}(b) shows time- and latitude-root-mean-square (rms), radial profiles of B$_{\phi}$ (blue) and B$_r$ (red). Generally, the radial and azimuthal magnetic field strengths are the largest in the convection zone decreasing steeply at the convective-radiative boundary. Furthermore, the toroidal field component can be seen to be larger than the radial field component inside most of the convection zone and the near-core region. At the CRB, B$_r$ = 33.8 $\pm$ 18.0 kG and B$_{\phi}$ = 120.8 $\pm$ 60.2 kG, whilst at the top of the \bvf{} peak, B$_r$ = 0.292 $\pm$ 0.290 kG and B$_{\phi}$ = 7.65 $\pm$ 9.06 kG. These averages are within a factor of a few of the minimum field strengths inferred for KIC 9244992 by \cite{masao2025}. Also, note that the "uncertainty" shown for the values above are standard deviations, so the larger-than-mean standard deviations are simply stating large dispersion of values. We show the ratio of B$_{\phi}$ to B$_{r}$ in Fig.~\ref{fig:vp-br-bp-r}(c), which shows that B$_{\phi}$ is generally larger than B$_{r}$ in the convection zone, but rises steeply and peaks within the \bvf{} frequency spike, with the maximum ratio of 20. 


An important point to address is how a percentage difference between the "core" rotation rate and surface rotation rate of 3.7\% from our simulations (see Fig.~\ref{fig:vp-br-bp-r}(a)) can lead to such large toroidal fields. The small percentage difference is calculated from the averaged radial rotation profile and we expect these averaged values to be closest to the minimum values inferred through asteroseismology. \cite{Saio2015} estimated that for KIC 9244992 rotation periods of the core and the envelope to be 63.51 $\pm$ 0.28 d and 66.18 $\pm$ 0.58 d, respectively, which are horizontal- and time-averaged values, which gives a percentage difference of 4.03\%. Notably though, the "core" value reported in \cite{Saio2015} is the {\it near-core} region, specifically the region contained within the BV-spike and not the convective core itself.  This is a crucial distinction.  The differential rotation between the {\it near-core} and surface measured by \cite{Saio2015} and that measured in our simulation are remarkably close, despite the difference in absolute rotation rate.  However, looking at the spread of the differential rotation at different latitudes in Fig.~\ref{fig:vp-br-bp-r}(a), we see that this can be large, especially closer to the equator, where the differential rotation can be as large as 20\%, which can effectively induce large toroidal fields, which is clearly observed in Fig.~\ref{fig:vp-br-bp-r}(b). 

To illustrate how much variation in rotation we observe across latitude, we show the latitudinal rotation profile in Fig.~\ref{fig:vp-br-bp-r-latdep}(a). These values are averaged in time and longitude, and the shaded region shows the variation of the profile across the \bvf{} frequency spike. The surface rotation rate is marked by the dotted, horizontal, black line. The profile indicates slower rotation at the equator compared to the mid-latitude and polar regions, characteristic of an anti-solar-type profile. Looking at the time- and longitude-averaged latitudinal rms toroidal field (blue lines) in Fig.~\ref{fig:vp-br-bp-r-latdep}(b) at different radii within the \bvf{} spike, the profiles show a double peak structure with the largest values around mid-latitudes and smallest at the poles. For comparison with values from \cite{masao2025} for KIC 9244992, we see the averaged field strengths at the peak of the \bvf{} spike to be close to the asteroseismically inferred values. This "radius for best comparison" is further supported by the ratio $B_{\phi,\mathrm{rms}}/B_{r,\mathrm{rms}}$, where we show in Fig.~\ref{fig:vp-br-bp-r-latdep}(c) that the ratios are more in line with the ratio of 30 given in \cite{masao2025}. Note that these ratios show the same dependence with latitude as that shown by $B_{\phi}$.

Finally, \citep{masao2025} describe the latitudinal geometry of the field by measuring separately the sign of the parameters a and b, which are related to the second order legendre polynomial (see their Equations 15-16), finding that a is < 0 and b is > 0.  We show these two functions in FIG.~\ref{fig:abomega}, where indeed the simulations agree with the asteroseismic inference on the broad latitudinal structure.  

\vspace{-0.9cm}
\section{Simulation Parameters}\label{sec:Simulation_Parameters}
As discussed in ~\ref{sec:Numerical_setup}, simulation parameters such as $Re$, $Rm$ and $Ro$ play a crucial role in the development of a dynamo generated field and in this paper, we presented what we view as the model that would match KIC 9244992 the best, in terms of these paramaters. To check the effect of different $Ro$, $Re$ and $Rm$, we ran two other simulations varying the rotation rate and convective velocities. In our second model, the rotation period was set to 63 d, similar to that inferred in \cite{masao2025}, whilst the convective forcing was reduced such that the convective velocities were 3--4 $\times$ MLT velocities, resulting in a $Re \sim 120$ and $Ro \sim$0.8.  This model produced magnetic field strengths and geometry similar to the model presented above (and KIC 9244992), but the latitudinal differential rotation was reduced. Therefore, this model too, would have been in agreement with the observations. In the third model, the rotation rate was again 63 d but the convective forcing was reduced to achieve convective velocities that are more similar to MLT velocities.  In this model, where $Re \sim 30$ and $Ro \sim 0.1$, the magnetic field is substantially reduced to only 1--2kG, inconsistent with the observations.  We expect that the weak field in this case is due to the very low Reynolds number, but running simulations with MLT velocities at high Reynolds numbers has proven numerically challenging in these global coupled simulations. 

Though this isn't a full parameter study, the results are consistent with previous simulations. For example, despite having a significantly higher ratio of ME/KE and lower Rossby number, the simulations of a 2M$_{\odot}$ star from \cite{Featherstone2009} give a similar field strength to what we've found here. Similarly, the simulations of a 10M$_{\odot}$ star from \cite{Auguston2016} have a large range of ME/KE and Rossby number and all give an rms magnetic field strength similar to what's found here. While theory may suggest a tight relationship between magnetic and kinetic energy, numerical simulations show a much broader scatter \citep{Auguston2019} indicating the force balances from which such scaling laws are derived are more complex than typically assumed.

 We have shown that provided that the Rossby number of the simulation is similar to the model reported in \citep{masao2025} for MLT velocities, and that the $Re$ is large enough that nonlinear and inductive terms are dominant over dissipative ones, the dynamo generated field is significant and the geometry is such that the toroidal field dominates the radial field at the CRB. 

Since our simulations are far more dissipative than the star, due to the larger magnetic resistivities used for numerical stability, it is worth determining whether the magnetic field in the \bvf{} frequency spike is just due to diffusion out of the convection zone, which would happen on a much faster timescale in our simulations than in the actual star. Over the timescale of the simulation, a large scale field would have diffused well beyond the width of the BV spike, {\it if} diffusion were dominant in this region. However, diffusion is {\it not} dominant in that region. Induction of toroidal field from the shear flow is actively generating toroidal field from the radial field. Moreover, advection of field due to convective overshoot is also occurring in this region, amplifying both poloidal and toroidal field in this region. The relative amplitude of the (dominant) induction term to diffusion can be obtained simply by comparing $B_r\partial{v_{\phi}}/{\partial{r}}$ with $\eta\partial^2{B_{\phi}}/\partial{r}^2$. Using the numbers in Fig.~\ref{fig:vp-br-bp-r} one can estimate that the inductive term is approximately 100 times the diffusive term, consistent with the magnetic Reynolds number in that region. Therefore, despite our larger than physical resistivity, the region is dominated by induction, as it would be in a star. Hence, the field recovered in the presented simulation is also likely reflective of the physics occurring in the actual star.   

In addition to the simulations described above, we performed a run with an initial magnetic field strength reduced by a factor of five. We find that the resulting statistically steady field strength within the convection zone and CRB is unchanged. The primary difference is that the lower-initial-field case requires a longer time to reach this saturated state.

\section{Conclusions} \label{sec:conclusion}
We have presented 3D anelastic MHD simulations of a rotating 2 \msol{} main-sequence star, aimed at testing whether the predominantly toroidal magnetic fields inferred asteroseismically for the F-type star KIC 9244992 can arise naturally in global stellar dynamo models. This work follows up on previous works such as \cite{Featherstone2009}, \cite{Auguston2016} and \cite{ratnasingam2024}.  Despite differences in mass, rotation rate and numerical parameters, all these simulations produce dynamo fields of the order 10$^5$ G, regardless of whether they are sub-equipartition, equipartition or super-equipartition, lending weight to this as a typical field strength in core-convecting stars.  Moreover, in all these simulations the toroidal field is (at least) equivalent in strength to the radial (poloidal) field within the convection zone, though the specific field geometry within adjacent stable layers is not generally stated. Although only \cite{ratnasingam2024} and the current work look specifically in the region accessible to asteroseismology, the combined works paint a consistent picture of the field strength and geometry in the near core region of core-convecting stars. 

Despite the higher rotation rate of our model compared to that inferred for KIC 9244992 by \cite{masao2025}, we find that the relative differential rotation between the near-core region and the surface is comparable to that inferred observationally. Localized shear layers near the convective–radiative boundary efficiently amplify the toroidal field through the $\Omega$-effect, producing azimuthal-to-radial field ratios of order 5 -- 20 in the near-core region. The resulting field strengths are within the same order of magnitude as the minimum values inferred for KIC 9244992, providing independent numerical support for the interpretation of the asteroseismic observations \citep{masao2025}. Furthermore, we present further proofs that our simulations agree with the inferences made in \cite{masao2025}, by showing that two inequalities derived in that work, which are $a < 0 $ and $b > 0$ (see eq. 6 in \cite{masao2025}), agree with our simulation results. We also calculated the horizontal average of the stellar period using the rotation kernel of dipole modes, similar to the methods used to infer rotation rates in asteroseismology. Comparing this profile with the simple spherical average adopted in this work, we find that the near-core rotation rate is approximately 7.7 days and the surface rotation rate is approximately 7.3 days, which falls within the shaded region of the period profile shown in Fig.~\ref{fig:vp-br-bp-r}(a). We explain these quantities in more detail in Appendix~\ref{ap:A}. Note that the quantity most relevant for comparison with asteroseismic measurements is the contrast between the near-core and surface rotation rates, rather than the detailed radial structure of the rotation profile.

Our analysis further shows that latitudinal variations in the toroidal field are closely tied to the latitudinal structure of differential rotation, with peak azimuthal field strengths occurring at mid-latitudes rather than being strictly confined to the equatorial regions. Most important, however, is how the local differential rotation induces large toroidal fields, but presents as uniform rotation when measured through asteroseismic methods. This work highlights the importance of three-dimensional shear in shaping internal magnetic field geometry and cautions against overly simplified assumptions of the magnetic field geometry in asteroseismic inversions.

\section*{Acknowledgements}
We acknowledge support from STFC grant ST/W001020/1. Computing was carried out on the DiRAC Data Intensive service at Leicester (DIaL), operated by the University of Leicester IT Services, which forms part of the STFC DiRAC HPC Facility (\url{www.dirac.ac.uk}), funded by BEIS capital funding via STFC capital grants ST/K000373/1 and ST/R002363/1 and STFC DiRAC Operations grant ST/R001014/1. The authors thank S. Hekker, M.~S. Cunha and M. Takata for their input on the observational context of the work.

\section*{Data Availability}
The data underlying this article will be shared on reasonable request to the corresponding authors.


\bibliographystyle{mnras}
\bibliography{paper5} 



\appendix
\section{Extra Observational and Simulation Validation}\label{ap:A}
\begin{figure}
        \centering
        \includegraphics[trim={0.cm 0.cm 0.cm 0.0cm},clip,width=\columnwidth]{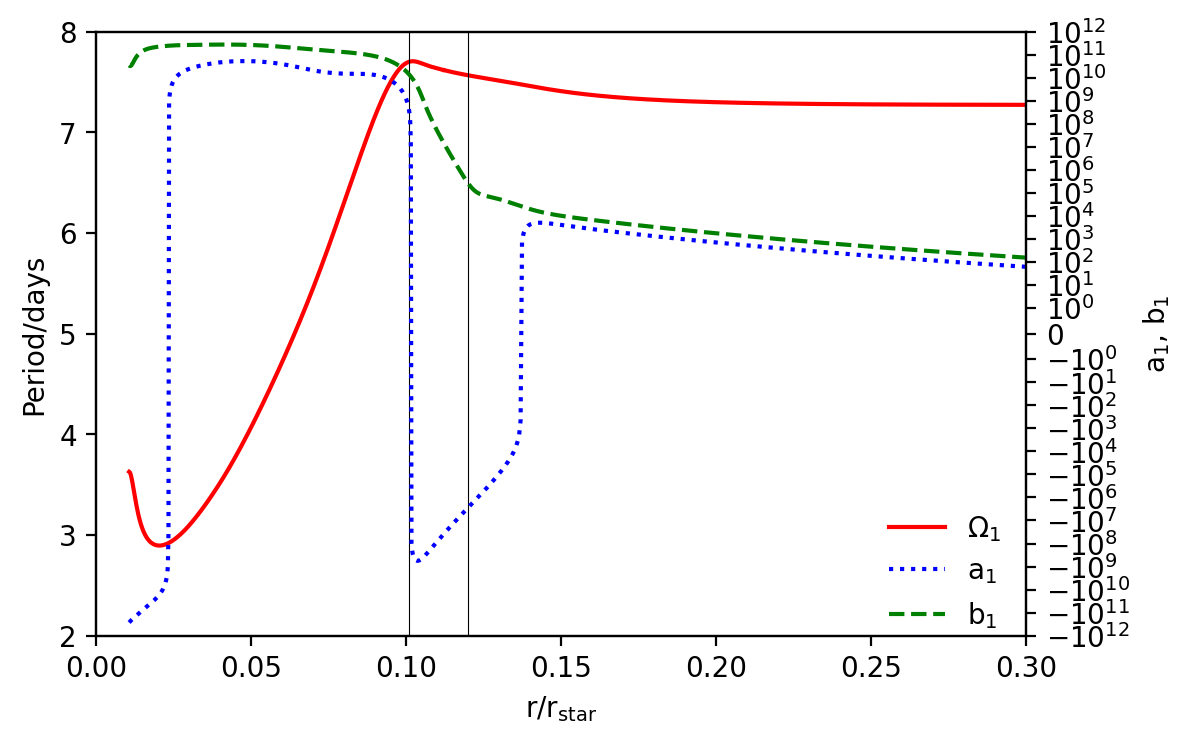}
        \caption{The rotation period (= 1 / $\Omega_1$), $a_1$ and $b_1$ as functions of radius \citep{masao2025}. \label{fig:abomega}}
        \centering
\end{figure}

In \cite{masao2025}, it was inferred that $a<0$ and $b>0$ (see equation 6 in \cite{masao2025}), which was used to make two simple and import remarks; the negative sign of $a$ implying
that the magnitude of the radial magnetic field is larger on average near the equatorial region than in the polar region and that the positive sign of $b$ means that azimuthal magnetic field is more confined to the equator than the poles. To show that these inequalities hold in our simulations, we can simplify them to the equations shown below:
\begin{equation}
a_1 = \int B_r^2\;( 3\cos^2 \theta - 1) \sin \theta\; d\theta\; d\phi < 0,   
\end{equation}
\begin{equation}
b_1 = \int B_\phi^2\;( 5\cos^2 \theta - 1) \sin \theta\; d\theta\; d\phi > 0.
\end{equation}
Furthermore, in terms of the asteroseismic measurement of the near-core rotation rate, the horizontal average of the rotation rate takes the following format:
\begin{equation}
\Omega_1 = \frac{3}{4} \int \Omega(r,\theta) \sin^2 \theta\; d\theta.   
\end{equation}
The quantity $\Omega_1$ represents the rotation rate inferred from dipole g-modes and therefore provides a more direct analogue to asteroseismic measurements than a simple spherical average. Because this measure involves a non-uniform latitudinal weighting, it is sensitive to latitudinal differential rotation, which is substantial in our simulations (Fig. \ref{fig:vp-br-bp-r-latdep}). This explains the differences in shape between Fig.~\ref{fig:abomega} and Fig.~\ref{fig:vp-br-bp-r}(a), despite similar near-core to surface contrasts.



\bsp	
\label{lastpage}
\end{document}